\documentclass[twocolumn,english]{revtex4-1}
\usepackage[T1]{fontenc}
\usepackage[latin9]{inputenc}
\usepackage[a4paper]{geometry}
\usepackage{amsmath}
\usepackage{amssymb}
\usepackage{esint}

\makeatletter

\usepackage{hyperref}

\makeatother

\usepackage{babel}
\begin{document}
\title{Asymmetrically twisted strings}
\author{Renann Lipinski Jusinskas}
\affiliation{Institute of Physics of the Czech Academy of Sciences \& CEICO ~\\
Na Slovance 2, 18221 Prague, Czech Republic}
\begin{abstract}
In this letter a new class of twisted strings is presented, with an
asymmetry between the holomorphic and antiholomorphic sectors parametrized
by an integer $N$. Their physical content is given by the massless
resonances of the closed string plus the mass-level $N$ spectrum
of the open string. The appeal of this model is the singling out of
the (higher spin) massive levels of string theory together with their
self/gauge/gravity interactions. Motivated by the original tree level
Kawai-Lewellen-Tye relation for closed strings, its asymmetrically
twisted version at four points is conjectured and shown to naturally
interpolate with conventional and twisted strings. The resulting four-point
amplitudes have a generalized Virasoro-Shapiro dressing factor.
\end{abstract}
\maketitle

\section{Overview}

Recent years have witnessed an unprecedented interplay between string
theory and quantum field theory. Perhaps the most evident example
of this mutual exchange is the Cachazo-He-Yuan (CHY) formalism \cite{Cachazo:2013iaa,Cachazo:2013hca,Cachazo:2013iea},
which describes field theory amplitudes through moduli space integrations
over a Riemann surface. Indeed, the CHY formulae were promptly shown
to emerge from a chiral worldsheet model: the ambitwistor string \cite{Mason:2013sva}. 

The connection between scattering amplitudes and the worldsheet should
not come as a surprise. After all, string theory has been exploring
it for about half a century. However, string amplitudes are in a sense
tainted by an infinite spectrum, with arbitrarily high mass levels
parametrized by the string length squared $\alpha'$. This is in contrast
to one striking feature of the ambitwistor string and the original
CHY formulation, which exclusively describe massless states.

Simultaneously to the debut of the scattering equations, Hohm, Siegel,
and Zwiebach introduced a string-like model with manifest $T$-duality
\cite{Hohm:2013jaa}. This model was later shown to have the ambitwistor
string as its tensionless limit \cite{Siegel:2015axg}. The tensionful
version is characterized by a certain twist between the left- and
right-moving sectors of the target space coordinates, hence the epithet
\emph{twisted }string (not to be confused with the twisted string
states in the context of orbifolds, e.g. \cite{Dixon:1986qv}). This
twist has a dramatic effect on the string spectrum. Besides the usual
massless states of the \emph{closed} string, the infinite tower of
massive states gives way to a single level, with the degrees of freedom
of the first massive level of the \emph{open} string. Spectrum wise,
it is fair to say that the twisted string lies between ambitwistor
and conventional strings.

While most of the current research in scattering amplitudes revolves
around massless fields, the twisted string incidentally let us hope
of a more general and controlled study of massive resonances. The
reason is precisely its finite spectrum, which can then be described
through a Lagrangian. This fact has been recently explored in \cite{Guillen:2021mwp},
introducing a surprisingly compact method for computing tree level
string amplitudes involving one massive multiplet at the first mass
level. This method is based on field theory techniques that shortcut
the traditional conformal field theory (CFT) computations intrinsic
to the string scattering.

The obvious question at this point is: what is special about the first
massive level of the string spectrum? From the string cohomology perspective,
not much. The twisted string is just a particular case of a deformation
of the target space CFT. This deformation is parametrized by a positive
integer $N$ which corresponds to the mass level it singles out. In
string theory, however, higher mass levels are tied to higher spins
(Regge trajectories). One is then bound to find non-trivial aspects
in the tree level amplitudes of such models, for they involve interactions
of higher spin states. Here I will discuss this construction and some
related implications.

\section{The asymmetric twist}

The asymmetric twist is defined as a deformation of the matter CFT,
encoded in the OPE of the target space coordinates $X^{m}$,
\begin{multline}
X^{m}(z,\bar{z})\,X^{n}(y,\bar{y})\sim-\tfrac{\alpha'}{2}\eta^{mn}\ln(z-y)\\
+\tfrac{\alpha'}{2N}\eta^{mn}\ln(\bar{z}-\bar{y}).\label{eq:XXOPE}
\end{multline}
Here, $\eta^{mn}$ is the flat space metric. The twisted string is
reproduced when $N=1$, while $N=-1$ takes us back to the conventional
string.

The left- and right-moving components of the $X^{m}$ energy-momentum
tensor are respectively given by
\begin{equation}
\begin{array}{ccc}
T=-\tfrac{1}{\alpha'}\partial X^{m}\partial X_{m}, &  & \bar{T}=\tfrac{N}{\alpha'}\bar{\partial}X^{m}\bar{\partial}X_{m},\end{array}
\end{equation}
such that the central charge contribution from the target space coordinates
does not depend on $N$.

Nevertheless, eigen-momentum operators now have different conformal
weight in the two sectors: $e^{ik\cdot X}$ (normal ordering implicit)
has conformal weight $\alpha'k^{2}/4$ in the holomorphic sector and
$-\alpha'k^{2}/(4N)$ in the antiholomorphic sector. Relatedly, the
Koba-Nielsen factor at tree level is expressed as
\begin{eqnarray}
\textrm{KN}_{N} & \equiv & \Big\langle\prod_{j=1}^{n}e^{ik_{j}\cdot X(z_{j},\bar{z}_{j})}\Big\rangle,\nonumber \\
 & = & \prod_{1\leq i<j}\Big(\frac{z_{ij}}{\bar{z}_{ij}^{1/N}}\Big)^{\alpha'(k_{i}\cdot k_{j})/2},\label{eq:Koba-Nielsen}
\end{eqnarray}
where $z_{ij}=z_{i}-z_{j}$. The structure of $\textrm{KN}_{N}$ partially
reflects the physical spectrum of the theory. For $N=-1$, the tree
level amplitudes computed using the Kawai-Lewellen-Tye (KLT) relation
\cite{Kawai:1985xq} have poles matching the infinite number of string
resonances. For the twisted string ($N=1$), a modified KLT construction
is required \cite{Huang:2016bdd,Mizera:2017cqs,Mizera:2017rqa}, leading
to a simplified pole structure matching its finite spectrum.

For $N>1$, the twist in (\ref{eq:XXOPE}) also leads to a finite
massive spectrum in bosonic and heterotic string theories. They will
be called here asymmetrically twisted strings (ATS).

\section{The physical spectrum}

The physical states of the ATS are given by the massless resonances
of conventional \emph{closed} strings and the mass level $N$ resonances
of the \emph{open} (super) string. For the sake of simplicity, I will
consider here the bosonic string and the spinning string with $\mathcal{N}=(1,0)$
worldsheet supersymmetry. The extension to heterotic string formalisms
with manifest spacetime supersymmetry follows almost effortlessly.

As in conventional string theories, the physical spectrum of the ATS
is given by the cohomology of the string BRST charge at ghost number
two. In addition, physical states are independently annihilated by
the zero mode of the reparametrization antighosts, $b$ and $\bar{b}$.
This is known as the Siegel gauge. The corresponding vertex operators
can then be cast as
\begin{equation}
U=c\bar{c}V_{L}\bar{V}_{R}e^{ik\cdot X},\label{eq:vertex}
\end{equation}
where $c$ and $\bar{c}$ are the reparametrization ghosts. $V_{L}$
and $\bar{V}_{R}$ are built using the remaining holomorphic and antiholomorphic
worldsheet variables. The vertex $U$ is independently annihilated
by the BRST charges $Q$ and $\bar{Q}$, which can be translated to
\begin{eqnarray}
Q\cdot(cV_{L}e^{ik\cdot X}) & = & 0,\label{eq:Q-eom}\\
\bar{Q}\cdot(\bar{c}\bar{V}_{R}e^{ik\cdot X}) & = & 0.\label{eq:Qbar-eom}
\end{eqnarray}
Since the BRST cohomology can only contain conformal weight $(0,0)$
vertex operators, $V_{L}$ and $\bar{V}_{R}$ are constrained to have
conformal weight $(1-\alpha'k^{2}/4,0)$ and $(0,1+\alpha'k^{2}/4N)$,
respectively.

\subsection*{Bosonic ATS}

The twisted bosonic string ($N=1$) has an extra $\mathbb{Z}_{2}$
symmetry of the type $\alpha'\to-\alpha'$ in its physical spectrum.
In addition to the usual massless states (graviton, dilaton and Kalb-Ramond
$2$-form), its physical content includes the first massive level
of the open string with $m^{2}=4/\alpha'$ and a mirrored tachyonic
state with $m^{2}=-4/\alpha'$. For $N>1$, this $\mathbb{Z}_{2}$
symmetry is absent and the physical spectrum has no tachyons.

The BRST charge is given by
\begin{equation}
Q=\oint(cT+bc\partial c),
\end{equation}
with a similar construction in the antiholomorphic sector using $\bar{T}$
and respective ghosts $(\bar{b},\bar{c})$. Nilpotency of $Q$ and
$\bar{Q}$ requires $D=26$. 

For positive integers $N$, there are only two simultaneous solutions
for $V_{L}$ and $\bar{V}_{R}$. The massless solution $k^{2}=0$
corresponds to $\bar{V}_{R}=\bar{a}_{m}\bar{\partial}X^{m}$, with
$k^{m}\bar{a}_{m}=0$ and $\bar{a}_{m}\cong\bar{a}_{m}+k_{m}\bar{\lambda}$.
In this case, the solution of (\ref{eq:Q-eom}) is given by $V_{L}=a_{m}\partial X^{m}$,
with $k^{m}a_{m}=0$ and $a_{m}\cong a_{m}+k_{m}\lambda$. This vertex
operator describes the massless excitations of the closed bosonic
string.

There is also a massive solution of (\ref{eq:Qbar-eom}) with $k^{2}=-4N/\alpha'$
and $\bar{V}_{R}=1$. Its left-moving counterpart in (\ref{eq:Q-eom})
matches the cohomology of the open bosonic string at mass level $N$.
In other words, the massive spectrum of the bosonic ATS has the mass
level $N$ degrees of freedom of the open string.

For $D<26$, it is still possible to build a critical bosonic string
via an affine algebra extension with currents $J^{A}$ and $\bar{J}^{a}$
for the holomorphic and antiholomorphic sector, respectively. The
massless spectrum is then enhanced by two gauge vectors, $\mathcal{A}_{m}^{A}$
and $A_{m}^{a}$, and a biadjoint scalar, $\phi^{Aa}$. In principle,
this setup would enable the extension of the techniques developed
in \cite{Guillen:2021mwp} for the computation of tree-level amplitudes
in conventional bosonic string theory. As will be clear soon, however,
a possible field theory description of the ATS spectrum with $N\geq2$
leads to a Lagrangian with an arbitrarily high number of derivatives.

\subsection*{Heterotic ATS}

For the heterotic theory, the analysis is very similar. $\bar{V}_{R}$
is built using $\bar{\partial}X^{m}$ and the current $\bar{J}^{a}$.
For $V_{L}$, the available ingredients are $\partial X^{m}$ and
its worldsheet superpartner $\psi^{m}$, and a $\beta\gamma$ system.
For more details, see e.g. \cite{Friedan:1985ge}.

The simultaneous solutions for $V_{L}$ and $\bar{V}_{R}$ can again
be divided into two groups. The massless spectrum is given by the
known $\mathcal{N}=1$ supergravity and super Yang-Mills states of
the conventional heterotic string. The massive solutions $k^{2}=-\tfrac{4N}{\alpha'}$,
with $\bar{V}_{R}=1$, match the open superstring spectrum with mass
level $N$ as can be readily seen by the output of (\ref{eq:Q-eom}).
Just like in the bosonic case, the dynamics of the heterotic ATS states
with $N\geq2$ cannot be completely captured by a finite Lagrangian.

\section{The ATS KLT matrix}

Tree level amplitudes in the asymmetrically twisted string are computed
using the standard prescription of the conventional string, with the
major difference given by the Koba-Nielsen factor in (\ref{eq:Koba-Nielsen}).

In the case of four external states, with (dimensionless) Mandelstam
variables $s_{ij}=\alpha'(k_{i}+k_{j})^{2}/4$, the ATS tree level
amplitudes can be expressed in terms of the following moduli space
integrals,
\begin{multline}
\ensuremath{\left.I_{N}\right._{(\bar{m},\bar{n})}^{(m,n)}}(s,t)=\\
\int d^{2}z\{z^{m+s}(1-z)^{n+t}\bar{z}^{\bar{m}-s/N}(1-\bar{z})^{\bar{n}-t/N}\},\label{eq:IN-integrals}
\end{multline}
where $\{m,n,\bar{m},\bar{n}\}$ are integer numbers, $s=s_{12}$,
$t=s_{13}$, and $u=s_{14}$. 

For a generic $N$, $I_{N}$ is not well defined in the complex plane
because of the branch cuts for non-integer $s$, $t$ and $u$. For
$N=1$ this problem is overcome using intersection theory \cite{Mizera:2017cqs,Mizera:2017rqa}.
For $N>1$, the only way to make sense of the ATS scattering amplitudes
is to attach to the prescription an extra input that effectively picks
the \emph{physical} branches when evaluating the integrals (\ref{eq:IN-integrals}).

One can anticipate some general features in the related tree level
amplitudes. Among them, gauge invariance (decoupling of BRST exact
states), crossing symmetry, and kinematic poles matching the mass
spectrum of the theory. Instead of looking for the precise definition
of the extra input to solve $I_{N}$, I will partially replicate the
original KLT analysis \cite{Kawai:1985xq} and propose a new KLT matrix.
This will lead to closed-form expression for $I_{N}$ that satisfies
the expected physical requirements and naturally interpolates with
the known cases $N=\pm1$.

Given the character of the branch cuts in (\ref{eq:IN-integrals}),
it is reasonable to expect the following behavior,
\begin{eqnarray}
I_{N} & \sim & Q(s,t)\times\Omega_{N},\label{eq:IN-expected}\\
\Omega_{N}(s,t,u) & \equiv & \tfrac{\Gamma(-s/N)}{\Gamma(-s)}\tfrac{\Gamma(-t/N)}{\Gamma(-t)}\tfrac{\Gamma(-u/N)}{\Gamma(-u)},\label{eq:dressing}
\end{eqnarray}
where $Q(s,t)$ is some rational function of the Mandelstam variables.
The argument to support the scalar dressing factor $\Omega_{N}$ is
threefold. First, it reproduces a Virasoro-Shapiro-like structure
for $N=-1$. Second, when $N=1$, the dressing factor collapses to
the unity, in agreement with the twisted string. Third, there are
no poles for $N>1$, which matches the finite character of the physical
spectrum of the ATS. In order to see this, recall that gamma functions
have no zeros, just poles. The possible poles in the dressing factor
come from positive $s$, $t$ or $u$ that are multiples of $N$.
In this case, both numerator and denominator are divergent, but their
ratio is well-defined and finite.

The KLT construction suggests the following output for the integrals
in (\ref{eq:IN-integrals}),
\begin{multline}
\ensuremath{\left.I_{N}\right._{(\bar{m},\bar{n})}^{(m,n)}}(s,t)=\int_{0}^{1}d\xi\,\xi^{m+s}(1-\xi)^{n+t}\\
\times(\textrm{KLT}_{N})\int_{1}^{\infty}d\eta\,\eta^{\bar{m}-s/N}(1-\eta)^{\bar{n}-t/N},\label{eq:IN-KLT}
\end{multline}
where $\textrm{KLT}_{N}$ carries the information about the physical
branches of $I_{N}$. There is then a simple solution for the modified
KLT matrix compatible with the above considerations:
\begin{equation}
\textrm{KLT}_{N}=\frac{\sin(\pi s)\sin(\pi t)}{\sin(\pi s/N)}.\label{eq:KLT-N}
\end{equation}

Using this result and Euler's reflection formula, $I_{N}$ can be
finally expressed as
\begin{multline}
\ensuremath{\left.I_{N}\right._{(\bar{m},\bar{n})}^{(m,n)}}(s,t)=\left[\frac{\sin(\pi s)}{\sin(\pi s/N)}\frac{\sin(\pi t)}{\sin(\pi t/N)}\right]\times\\
\frac{\Gamma(s+m+1)\Gamma(t+n+1)}{\Gamma(s+t+m+n+2)}\frac{\Gamma(\tfrac{s+t}{N}-\bar{m}-\bar{n}-1)}{\Gamma(\tfrac{s}{N}-\bar{m})\Gamma(\tfrac{t}{N}-\bar{n})}.\label{eq:IN-full}
\end{multline}
The term inside the square brackets essentially kills any simple pole
that is not a multiple of the level $N$. In addition, equation (\ref{eq:IN-full})
neatly recovers the conventional and twisted string results. One might
worry that poles at $2N,3N,\ldots$ might appear in the gamma functions,
which would explicitly spoil unitarity. However, the values of $\{m,n,\bar{m},\bar{n}\}$
at a given mass level are constrained by the conformal weight of the
chiral blocks $V_{L},\bar{V}_{R}$. It is then straightforward to
check that the ATS amplitudes obtained through (\ref{eq:IN-full})
cannot develop poles in $s,t,u$ greater than $N$.

\section{Some examples}

As a practical example of the conjectured moduli space integrals (\ref{eq:IN-full}),
we can look at the four-point amplitude with external gluons. It has
an universal character, offering several consistency checks. A straightforward
computation in the heterotic ATS yields
\begin{equation}
\mathcal{A}_{4}=\Omega_{N}(\mathcal{A}_{4}^{\textrm{SYM}}+\mathcal{A}_{4}^{\textrm{double}}),\label{eq:A4}
\end{equation}
where $\mathcal{A}_{4}^{\textrm{SYM}}$ is the field-theory four gluon
amplitude in super Yang-Mills and $\mathcal{A}_{4}^{\textrm{double}}$
encodes the double-trace contributions, given by 
\begin{multline}
\mathcal{A}_{4}^{\textrm{SYM}}=\epsilon_{1m}^{a}\epsilon_{2n}^{b}\epsilon_{3p}^{c}\epsilon_{4q}^{d}T^{mnpq}[\frac{(f_{abe}f_{ecd}-f_{cae}f_{ebd})}{st}\\
+\frac{(f_{bce}f_{ead}-f_{abe}f_{ecd})}{su}+\frac{(f_{cae}f_{ebd}-f_{bce}f_{ead})}{tu}],\label{eq:A4SYM}
\end{multline}
and
\begin{multline}
\mathcal{A}_{4}^{\textrm{double}}=3\kappa\epsilon_{1m}^{a}\epsilon_{2n}^{b}\epsilon_{3p}^{c}\epsilon_{4q}^{d}T^{mnpq}\\
\times(\frac{\delta_{ab}\delta_{cd}}{s(s+N)}+\frac{\delta_{ac}\delta_{bd}}{t(t+N)}+\frac{\delta_{ad}\delta_{bc}}{u(u+N)}),\label{eq:A4-double}
\end{multline}
where
\begin{multline}
T^{mnpq}=\eta^{mn}(uk_{1}^{p}k_{2}^{q}+tk_{2}^{p}k_{1}^{q})+\eta^{pq}(uk_{3}^{m}k_{4}^{n}+tk_{4}^{m}k_{3}^{n})\\
+\eta^{mp}(sk_{3}^{n}k_{1}^{q}+uk_{1}^{n}k_{3}^{q})+\eta^{nq}(sk_{4}^{m}k_{2}^{p}+uk_{2}^{m}k_{4}^{p})\\
+\eta^{np}(sk_{3}^{m}k_{2}^{q}+tk_{2}^{m}k_{3}^{q})+\eta^{mq}(sk_{4}^{n}k_{1}^{p}+tk_{1}^{n}k_{4}^{p})\\
+\tfrac{2}{\alpha'}(\eta^{mn}\eta^{pq}tu+\eta^{mq}\eta^{np}st+\eta^{mp}\eta^{nq}su).
\end{multline}
The gluon polarizations are denoted by $\epsilon_{m}^{a}$, while
$f_{abc}$ are the structure constants of the gauge group with level
$\kappa$. The amplitude (\ref{eq:A4}) has the ATS dressing factor
of (\ref{eq:dressing}), and the correct factorization channels. In
particular, the double-trace channels match the remaining spectrum
of the heterotic ATS, with contributions from the massless sector
($\mathcal{N}=1$ supergravity) and the mass level $N$ states of
the open superstring. 

When $N=1$, the dressing factor in (\ref{eq:A4}) goes away, leaving
behind the field theory amplitude characteristic of the twisted string.
For $N=-1$, the apparent tachyonic poles in (\ref{eq:A4-double})
are canceled out by the corresponding zero in $\Omega_{N}$. In this
case, (\ref{eq:A4}) reproduces the four gluon amplitude of the conventional
heterotic string. Observe that the residue of (\ref{eq:A4}) in the
massive pole (e.g. $s+N=0$) has a remarkably simple form,

\begin{multline}
\textrm{Res}\left.\mathcal{A}_{4}\right|_{s=-N}\propto\tfrac{\sin\pi t}{\sin\pi(t/N)}\times\\
\times\big\{(\epsilon_{1}^{a}\cdot\epsilon_{2}^{a})(\epsilon_{3}^{b}\cdot\epsilon_{4}^{b})\prod_{n=0}^{N}(t-N+n)+\ldots\big\}.\label{eq:A4res}
\end{multline}
Here the ellipsis denote other linearly independent contributions.
For $N=\pm1$ the prefactor disappears, the residue is a polynomial
in $t$ with degree $N+1$, and unitarity is granted. For generic
$N$, however, this is not the case. The prefactor in (\ref{eq:A4res})
implies an infinite expansion in any polynomial basis. This is a generic
behaviour of the ATS amplitudes, an imemdiate consequence of the dressing
factor $\Omega_{N}$.

For the bosonic ATS, I will quickly discuss a four-point amplitude
with $N=2$, involving a massive state and three gluons. The massive
resonance has spin $S=3$, described by the vertex operator $V_{L}=\phi_{mnp}\partial X^{m}\partial X^{n}\partial X^{p}$.
The totally symmetric polarization $\phi_{mnp}$ is traceless ($\eta^{mn}\phi_{mnp}=0$),
and transversal ($k^{m}\phi_{mnp}=0$). Its tree level scattering
with three gluons can be cast as
\begin{multline}
\mathcal{A}_{4}(\phi_{1},\epsilon_{2},\epsilon_{3},\epsilon_{4})\propto f_{abc}\phi_{mnp}^{1}\epsilon_{2q}^{a}\epsilon_{3r}^{b}\epsilon_{4s}^{c}\\
\times\Omega_{2}\Big\{\tfrac{4}{\alpha'}\eta^{ms}\eta^{nr}\eta^{pq}-2\eta^{nq}\eta^{pr}(k_{2}^{m}k_{2}^{s}+k_{3}^{m}k_{3}^{s})\\
-2\eta^{nr}\eta^{ps}(k_{3}^{m}k_{3}^{q}+k_{4}^{m}k_{4}^{q})-2\eta^{ns}\eta^{pq}(k_{2}^{m}k_{2}^{r}+k_{4}^{m}k_{4}^{r})\\
+\tfrac{2}{s}\eta^{nq}k_{2}^{m}[\eta^{pr}k_{3}^{s}(t+1)+\eta^{ps}k_{4}^{r}(u+1)]\\
-\tfrac{1}{s}\eta^{pq}\eta^{rs}[k_{3}^{m}k_{3}^{n}(u+1)+k_{4}^{m}k_{4}^{n}(t+1)]\\
+\tfrac{2}{t}\eta^{nr}k_{3}^{m}[\eta^{pq}k_{2}^{s}(s+1)+\eta^{ps}k_{4}^{q}(u+1)]\\
-\tfrac{1}{t}\eta^{pr}\eta^{qs}[k_{2}^{m}k_{2}^{n}(u+1)+k_{4}^{m}k_{4}^{n}(s+1)]\\
+\tfrac{2}{u}\eta^{ns}k_{4}^{m}[\eta^{pq}k_{2}^{r}(s+1)+\eta^{pr}k_{3}^{q}(t+1)]\\
-\tfrac{1}{u}\eta^{ps}\eta^{qr}[k_{2}^{m}k_{2}^{n}(t+1)+k_{3}^{m}k_{3}^{n}(s+1)]\\
+\tfrac{(t+1)(u+1)}{s(s-1)}\eta^{pq}\eta^{rs}k_{2}^{m}k_{2}^{n}+\tfrac{(s+1)(u+1)}{t(t-1)}\eta^{pr}\eta^{qs}k_{3}^{m}k_{3}^{n}\\
+\tfrac{(s+1)(t+1)}{u(u-1)}\eta^{ps}\eta^{qr}k_{4}^{m}k_{4}^{n}+\ldots\Big\}.
\end{multline}
The ellipsis inside the curly brackets consists of terms of the form
$(\alpha'k^{4})$ and $(\alpha'k^{3})^{2}$. Because of the color
structure, this amplitude has only massless poles, corresponding to
the exchange of gluons. The apparent poles at $s,t,u=1$ are also
canceled out here by the respective roots of the dressing factor.

In the heterotic ATS with arbitrary $N$, four-point amplitudes with
one massive leg at the leading Regge trajectory and three massless
ones can be determined using the very general construction of Schlotterer
in \cite{Schlotterer:2010kk}. This is possible because the CFT correlators
in the holomorphic sector of the ATS are essentially the same as in
the open superstring.

\section{Discussion}

The introduction of a parameter $N>1$ singling out a unique mass
level of the open (super) string spectrum from a worldsheet theory
is certainly enthralling. With a finite physical spectrum, this worldsheet
model may open the doors for a more systematic study of the massive
resonances of conventional string theory using field theory methods.
Especially appealing is the fact that the physical content of the
mass level $N$ contains states with spin $S=N+1$.

Higher spin field theories are well-known for their intricate dynamics,
severely constrained by a series of consistency requirements and no-go
theorems (see, for example, the review \cite{Rahman:2012thy} and
references therein). This behavior is supported by the four-point
amplitudes discussed here. They encode interactions involving an arbitrarily
high number of derivatives between massive higher spin fields, gauge
fields and gravity, manifested through the dressing factor $\Omega_{N}$.
Unlike the $N=1$ case, that can be exactly described by a finite
Lagrangian \cite{Guillen:2021mwp}, the dynamics of the ATS states
can only be captured by an effective field theory (EFT) through an
$\alpha'$-expansion. It would be really interesting to analyze such
EFTs more deeply, in particular in respect to the consistency of the
higher-spin interactions (for example, along the lines of \cite{Caron-Huot:2016icg}).
Unitarity is broken, cf. equation (\ref{eq:A4res}), unless there
is some nontrivial mechanism involving the infinite number of derivatives
of the EFT description. This analysis is left for future work. The
role of the $\textrm{KLT}_{N}$ matrix is particularly intriguing,
for it seems to connect an open string amplitude and what resembles
a chiral half of a string theory on an orbifold \cite{Dixon:1986qv}.

The analysis of the high energy behaviour of the integrals $I_{N}$
is straightforward, in particular the Regge limit and the hard scattering
limit. When we consider the scattering $1+2\to3+4$ of four states
with mass $m^{2}=4N/\alpha'$, and the center of mass reference frame
of the states $1$ and $2$, the Mandelstam variables read
\begin{equation}
\begin{array}{rcl}
s & = & -\tfrac{\alpha'}{4}E^{2},\\
t & = & (\tfrac{\alpha'}{4}E^{2}-4N)\sin^{2}\tfrac{\theta}{2},\\
u & = & (\tfrac{\alpha'}{4}E^{2}-4N)\cos^{2}\tfrac{\theta}{2},
\end{array}
\end{equation}
where $E$ is the center of mass energy and $\theta$ is the scattering
angle between states $1$ and $3$. The high energy behaviour ($\alpha'E^{2}\gg1$)
of the amplitudes is mainly driven by their dressing factors (\ref{eq:dressing}),
which in the above configuration can be expressed as
\begin{itemize}
\item Regge limit ($s\to-\infty$, small $\theta$):
\begin{equation}
\Omega_{N}\approx\tfrac{\Gamma(-t/N)}{\Gamma(-t)}s^{-(t+4N)(1-1/N)}.
\end{equation}
\item Hard scattering limit ($s\to-\infty$ and $s/t$ fixed):
\begin{equation}
\Omega_{N}\approx e^{(1-1/N)(s\ln s+t\ln t+u\ln u)}.
\end{equation}
\end{itemize}
Notice first that the usual Regge limit resonances for integer $t$
are absent, since the spectrum is finite. Second, the hard scattering
limit presents a soft string-like behaviour. This can again be explained
by the arbitrary number of higher derivatives of the effective field
theory for the higher spin fields.

At the level of the spectrum, the asymmetric twist has no impact on
type II string theories, either with $\mathcal{N}=(1,1)$ worldsheet
supersymmetry or explicit $\mathcal{N}=2$ spacetime supersymmetry:
there are only massless resonances. Intuitively, this can be explained
by the fact that the twist in (\ref{eq:XXOPE}) emulates a tachyon
vertex operator in the right-moving sector, which is not compatible
with the inbuilt supersymmetry.

Alternatively, when $N$ is taken to be a negative integer, the physical
spectrum again becomes infinite, albeit with an interesting change.
The level matching condition leads to an asymmetrical contribution
from the left- and right-moving parts of the vertex operators. Putting
it differently, the BRST cohomology is no longer comprised of worldsheet
scalars, unlike conventional strings. The chiral pieces $cV_{L}$
and $\bar{c}\bar{V}_{R}$ are similar to open (super) string vertex
operators but at different mass levels. The scope of this construction
can be extended by letting $N$ take rational values as well.

Differently from the $N=1$ case, the asymmetric twist is not obviously
connected to a singular gauge fixing of the Polyakov action \cite{Siegel:2015axg}.
Understanding the underlying gauge fixing leading to (\ref{eq:XXOPE})
might help to clarify the role of the parameter $N$ and to reveal
further applications of this model.

Note also that the twisted string ($N=1$) has a natural description
in terms of a chiral worldsheet \cite{Siegel:2015axg,LipinskiJusinskas:2019cej}.
The analysis of the $\alpha'\to\infty$ (tensionless) limit yielding
the ambitwistor string is very simple \cite{Bandos:2014lja,Casali:2016atr,Azevedo:2017yjy}.
The massive states in the bosonic and heterotic cases play the role
of auxiliary fields helping to implement higher derivative equations
of motion for the massless fields \cite{Berkovits:2018jvm,Azevedo:2019zbn}.
For $N>1$, there does not seem to exist a straightforward extension
of the chiral map of \cite{LipinskiJusinskas:2019cej}, and the tensionless
limit, if sensible, likely becomes more involved. In this case, the
higher spin states at mass level $N$ become massless and this investigation
might shed some light on the construction of ambitwistor strings for
higher spin fields.

For $N\to\infty$, the OPE (\ref{eq:XXOPE}) naively resembles that
of an open string. Indeed, the right-moving sector of the ATS becomes
inert. This can be seen through the BRST charge $\bar{Q}$. After
a field redefinition $\bar{c}\to\bar{c}/N$ and $\bar{b}\to N\bar{b}$,
it can be recast as
\begin{equation}
\bar{Q}=\tfrac{1}{\alpha'}\oint\,\bar{c}\bar{\partial}X^{m}\bar{\partial}X_{m}+\mathcal{O}(1/N).
\end{equation}
In the $N\to\infty$ limit, only the first term survives. Since the
OPE $\bar{\partial}X^{m}(z)X^{n}(y,\bar{y})$ is now regular, the
solutions of (\ref{eq:Qbar-eom}) become degenerate. Curiously, the
BRST cohomology at ghost number one, given in (\ref{eq:Q-eom}), is
enhanced and matches the physical degrees of freedom of the open (super)
string. It is enticing to combine this analysis with the tensionless
limit, as it hints at a possible route to define an \emph{open} ambitwistor
string.

Finally, a comment on string loops, i.e. higher genus surfaces. The
twisted string is not modular invariant \cite{Yu:2017bpw,Lee:2017utr}.
On the torus, with modular parameter $\tau=\tau_{1}+i\tau_{2}$, a
simple way to see this is to notice that the zero mode contribution
to the scalar partition function of $X^{m}$ is a function of $\tau_{1}$
instead of $\tau_{2}$. This is related to the sign flip in (\ref{eq:XXOPE})
for $N=1$. The asymmetric twist leads to a similar structure, with
a nontrivial mixing of the real and imaginary parts of the modular
parameter, which naively does not lead to a modular invariant partition
function. However, the loop level construction of the ATS can only
be properly done after the gauge structure of the worldsheet model
is understood.
\begin{acknowledgments}
I would like to thank Subhroneel Chakrabarti, Ted Erler, Henrik Johansson,
Sebastian Mizera and, in particular, Humberto Gomez and Oliver Schlotterer,
for valuable discussions, suggestions, and comments on the draft.
This research has been supported by the Czech Science Foundation -
GA\v{C}R, project 19-06342Y.
\end{acknowledgments}

\end{document}